\documentclass[prx,aps,amsmath,amssymb,showpacs,twocolumn]{revtex4-1}
\usepackage{amssymb,graphicx,float,dcolumn,bm,subfigure,color}
\usepackage[titletoc,title]{appendix}
\usepackage[dotinlabels]{titletoc}
\usepackage{cancel}
\usepackage{subfigure}
\usepackage{mathrsfs}
\usepackage[mathscr]{euscript}
\usepackage{hyperref}
\usepackage[normalem]{ulem}
\usepackage{hhline}

\usepackage{multirow, makecell, ragged2e}

\newcommand{\be}{\begin{equation}}
\newcommand{\ee}{\end{equation}}

\newcommand{\ba}{\begin{eqnarray}}
\newcommand{\ea}{\end{eqnarray}}

\renewcommand{\vec}[1]{\mbox{\boldmath$#1$}}

\newcommand\ie{\emph{i.e.}~}

\newcommand{\sh}{\mathcal{S}}

\def\beq{\begin{eqnarray}}

\newcommand{\PLLL}{{P_{\rm LLL}}}
\def\eeq{\end{eqnarray}}
\renewcommand{\i}{i}

\newcommand{\elliptic}[5][\scriptstyle]{\vartheta\left[\begin{array}{c}{{#1 #2}}\\{#1 #3}\end{array}\right]\left(#4\middle|#5\right)}

\makeatletter
\newcommand*{\rom}[1]{\expandafter\@slowromancap\romannumeral #1@}
\makeatother

\begin{document}

\title{Hall viscosity of the composite-fermion Fermi seas for fermions and bosons}

\author{Songyang Pu}
\affiliation{Department of Physics, 104 Davey Lab, Pennsylvania State University, University Park, Pennsylvania 16802, USA}

\date{\today}

\begin{abstract} 
The Hall viscosity has been proposed as a topological property of incompressible fractional quantum Hall states and can be evaluated as Berry curvature. 
This paper reports on the Hall viscosities of composite-fermion Fermi seas at $\nu=1/m$, where $m$ is even for fermions and odd for bosons. A well-defined value for the Hall viscosity is not obtained by viewing the $1/m$ composite-fermion Fermi seas as the $n\rightarrow \infty$ limit of the Jain $\nu=n/(nm\pm 1)$ states, whose Hall viscosities $(\pm n+m)\hbar \rho/4$ ($\rho$ is the two-dimensional density) approach $\pm \infty$ in the limit $n\rightarrow \infty$. A direct calculation shows that the Hall viscosities of the composite-fermion Fermi sea states are finite, and also relatively stable with system size variation, although they are not topologically quantized in the entire $\tau$ space. I find that the $\nu=1/2$ composite-fermion Fermi sea wave function for a square torus yields a Hall viscosity that is expected from particle-hole symmetry and is also consistent with the orbital spin of $1/2$ for Dirac composite fermions. I compare my numerical results with some theoretical conjectures.
\end{abstract}

\maketitle
\section{Introduction}
The Hall viscosity has been proposed as one of the topological characteristics of fractional quantum Hall (FQH) states~\cite{Avron95,Tokatly07b,Tokatly09,Read09}. In particular, it has been proposed that it is related to the ``shift" on the spherical geometry~\cite{Read09}, as confirmed by explicit evaluations~\cite{Fremling14,Pu20} of the Laughlin and Jain states~\cite{Laughlin83,Jain89}. (The shift is twice the so-called orbital spin \cite{Wen92}.) This paper is concerned with the Hall viscosity of the composite-fermion Fermi sea (CFFS). Composite fermions (CFs) form a Fermi sea when they experience a zero effective field~\cite{Halperin93,Jain07,Halperin20b,Shayegan20}. The best studied CFFS is at $\nu=1/2$, where electrons capture two vortices to form composite fermions. I will also consider the CFFSs of fermions at $\nu=1/4$ and $\nu=1/6$, where composite fermions bind four and six vortices. Just as fermions in the lowest Landau level capture an even number of vortices to form FQH states and the CFFSs, bosons in the lowest Landau level can capture an odd number of vortices to form both FQH states and the CFFSs~\cite{Cooper99,Regnault03,Chang05b,Wu13,Wu15b}. I will consider CFFSs of bosons at $\nu=1$, 1/3, and 1/5. 

A fundamental difficulty for the determination of the Hall viscosity of the CFFS is that it does not have a gap in the thermodynamic limit, and its Hall viscosity is not expected to be topologically quantized and may be sensitive to various details, such as the geometry of the torus, the shape and size of the CFFS, and the details of the CFFS wave function. Nonetheless, irrespective of the issue of its applicability to real experiments, the Hall viscosity can be evaluated for the standard CFFS wave functions, which are very accurate representations of the actual Coulomb ground states. This article reports on these results. I evaluate the Hall viscosity through calculation Berry curvature in the $\tau$ space following Avron, Seiler, and Zograf~\cite{Avron95} (explained below); the method is justified by the presence of a gap at individual $\tau$ points for a finite system. To this end, I construct CFFS wave functions for these states and show that for the general $1/m$ filling several wave functions can be constructed. 

For $\nu=1/2$ particle-hole symmetry is an additional consideration that fixes the value of the Hall viscosity, as shown by Read and Rezayi \cite{Read10}; I find that the calculated value is consistent with the expected value, which is not surprising given that the wave function satisfies particle-hole symmetry to a good degree. The value of Hall viscosity at $\nu=1/2$ is consistent with the orbital spin $1/2$ for Dirac composite fermions \cite{Son15,Levin17}, to the extent that the conjectured relation between the Hall viscosity and the orbital spin holds for incompressible states. I also show the Hall viscosities for CFFSs are not topologically quantized in the $\tau$ space, in stark contrast to the gapped FQH states. 

This paper is organized as follows. I first briefly review the Hall viscosity for gapped FQH states and the problem for CFFSs in Sec.~\ref{definition}. Then I introduce the wave functions for both fermionic CFFSs and bosonic CFFSs in Sec.~\ref{wf}. Finally, I present my results and discussions of Hall viscosity for CFFSs in Sec.~\ref{results}.

\section{Hall viscosity as Berry curvature}
\label{definition}
The Hall viscosity is a bulk property of quantum Hall fluid. It is the geometrical response to the strain rate applied to the fluid. In theoretical calculations, the strain rate can be simulated by putting the fluid on a torus and adiabatically deforming the shape of the torus while preserving its area. A torus is equivalent to a parallelogram on a complex plane with periodic boundary conditions in both directions, $L_1$ along the real axis and $L_2=L_1\tau$, where the modular parameter $\tau=\tau_1+\i \tau_2$ is a complex number \cite{Gunning62}. The total area is given by $V=L_1^2\tau_2=2\pi N_\phi \ell^2$, through which there are $N_\phi$ flux quanta passing. (A flux quantum is defined as $\phi_0=h/e$, and the magnetic length is defined as $\ell=\sqrt{\hbar c/eB}$.) It was shown by Avron, Seiler, and Zograf \cite{Avron95} that the Hall viscosity can be computed as Berry curvature through adiabatic deformation of the geometry of the torus: 
\be
\label{berry-curv}
\eta^A=-{\hbar \tau_2^2  \over V}\mathcal{F}_{\tau_1,\tau_2},
\ee
where 
\be
\label{BC}
\mathcal{F}_{\tau_1,\tau_2}=-2{\rm Im}\bigg\langle {\partial \Psi \over \partial \tau_1}\bigg|{\partial \Psi \over \partial \tau_2}\bigg\rangle.
\ee
Here $\Psi$ is the many-particle ground state on the torus. Based on Eq.~\ref{berry-curv}, Read proposed that $\eta^A$ is given by
\be
\label{hall visc}
\eta^A=\sh{\hbar \over 4}{N \over V}.
\ee
where the ``shift" $\sh$ is a topological quantum number, given by $\sh={N\over \nu}-N_\phi$; that is, $\sh$ is the offset of flux quanta needed to form a ground state with $N$ particles on a sphere \cite{Read09,Read10}.

The shift is a manifestation of the orbital spin \cite{Wen92} which is given by $\sh/2$.  Relation Eq.~\ref{hall visc} has been proved for various incompressible FQH states with methods of plasma analogy, Chern-Simons theory, matrix models {\em etc.} \cite{Read09,Read10,Tokatly09,Cho14,Lapa18,Lapa18b}. Most recently, Pu, Fremling, and Jain \cite{Pu20} found this relation can be proved for $\nu={m\over 2pm\pm 1}$ by using the Jain wave functions \cite{Pu17} and certain natural and justified assumptions.

\begin{figure}[t]
	\includegraphics[width=\columnwidth]{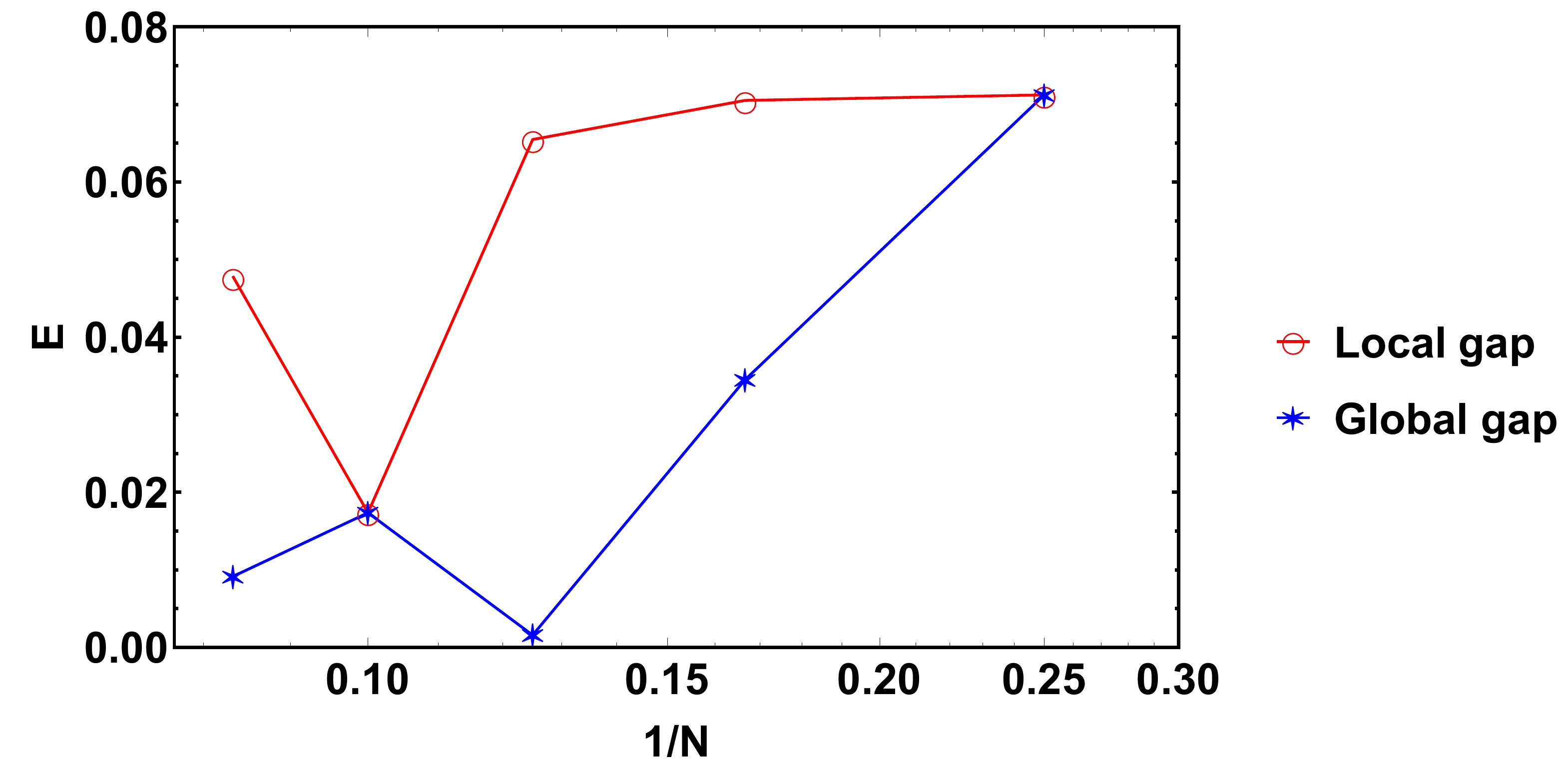} 
	\caption{Comparison of local gaps and global gaps obtained from exact diagonalization for a rectangular torus with $\tau=1.1\i$. The ``local gap'' is defined as the energy difference between the two lowest-energy states in the momentum sector that the ground state belongs to, and the ``global gap'' is defined as the energy difference between the two lowest-energy states in the Haldane momentum space. The Berry curvature is related to only the local gap since the momentum sector is fixed. The unit of energy is $e^2/\epsilon \ell$, and the x-axis is one over particle number. }
	\label{fig1}
\end{figure}

In this article, I am concerned with the Hall viscosity of gapless states. A large group of gapless states in FQH systems are compressible Fermi-liquid like states, which are described as composite-fermion Fermi seas. As the derivation of Eq.~\ref{berry-curv} is based on adiabatic transformations, one may question if it can be used to characterize the Hall viscosity for gapless states. Here, it is important to note that the Hall viscosity is defined as a Berry curvature, which requires a finite gap only in the vicinity of a certain value of $\tau$ \cite{Milovanovic10}. (In contrast, the Berry {\it phase} would require integration over the entire $\tau$ space.) One may argue that there is a finite gap for the individual value of $\tau$ for the compressible Fermi-liquid like states of finite size, even though there may be level crossings as a function of $\tau$ and for certain values of $\tau$ the chosen wave function may no longer represent the ground state accurately. 
Second, the Berry curvature is calculated within the ground state momentum sector. This means single particle-hole pair excitation is irrelevant because it changes the total momentum, and the relevant lowest-energy excitation is the creation of two particle-hole pairs of opposite momenta. In other words, the relevant gap for the Hall viscosity is the gap within a momentum sector that the ground state belongs to, which I call the ``local gap'', rather than the energy difference between the two lowest energy states in the whole Haldane momentum space \cite{Haldane85b} (``global gap''). In Fig.~\ref{fig1} I show a comparison of local gaps and global gaps obtained by exact diagonalization. While the system sizes calculated are quite limited, the tendency clearly indicates that the local gap is much bigger than the global gap in general and is present at least for finite systems.
Therefore, the Berry curvature in Eq.~\ref{berry-curv} is still well defined for a given model state, provided it corresponds to the ground state for the value of $\tau$ under consideration. 

One may think that the Hall viscosity of the CFFS at $\nu=1/m$ may be trivially obtained by viewing it as the $n\rightarrow \infty$ limit of the Jain states $\nu=n/(mn\pm 1)$. For the Jain states, the Hall viscosity is given by Eq. \ref{hall visc} with $\sh=\pm n+m$. This would imply that the Hall viscosity of $\nu=1/m$ is $+\infty$ along the sequence $\nu=n/(mn+1)$ and $-\infty$ along the sequence $\nu=n/(mn-1)$. This obviously leads to a contradiction. I find below a finite Hall viscosity by a calculation directly at $\nu=1/m$. Nonetheless, the question of how to reconcile the Hall viscosity of the $\nu=1/m$ CFFS with the Hall viscosity of nearby Jain states remains an interesting open issue.

Ref.~\cite{Read10} showed that the Hall viscosity for a particle-hole symmetric state would be ${\hbar N\over 4V}$ whether the model state is incompressible or not. This result is applicable to the CFFS at $\nu=1/2$. Although it was not explicit in the construction of the CFFS wave function, it was shown that the CFFS does preserve a high degree of particle-hole symmetry \cite{Balram17b,Geraedts18,Wang19,Fremling18}, which is an exact symmetry at $\nu=1/2$ in the limit of zero Landau level (LL) mixing for any two-body interaction.  My calculation below is consistent with this result. It is known that at $\nu=1/2$ the Coulomb ground state is well described by such a CFFS \cite{Halperin93,Willett93,Kalmeyer92,Rezayi94}. 
For CFFSs at other fillings there is no rigorous theoretical derivation of Hall viscosities, to my knowledge. It is not clear whether the Hall viscosities for these states are universal and how they are related to the orbital spin or the shift \cite{You16}. Based on the idea of attaching fluxes to Dirac composite fermion, Goldman and Fradkin \cite{Goldman18} made a conjecture about the values of Hall viscosities of general CFFSs, which are compared to my numerical results in Sec.~\ref{results}.

In the remaining part of this paper, I first introduce the generic wave functions for CFFS. I then numerically calculate the Berry curvature for the CFFS at filling $1/2$, which has a high degree (although not exact) of particle-hole symmetry, and also CFFSs at other fillings, including $1, 1/3, 1/4, 1/5$, and $1/6$. 

\section{Composite fermion Fermi sea}
\label{wf}

I consider a parallelogram with edges $L_1$ and $L_2=L_1\tau$. Because the physical coordinate $z_i=x_i+iy_i$ of a particle changes with the deformation of the torus, it is convenient to define the reduced coordinate $(\theta_{1,i},\theta_{2,i})$ as $z_i=x_i+iy_i=L_1\theta_{1,i}+L_2\theta_{2,i}$. The reduced coordinates $(\theta_{1,i},\theta_{2,i})$ remain unchanged under $\tau$ deformations, and are therefore 
more convenient to use in the calculation of the Hall viscosity. Following the convention in Ref.~\cite{Pu20,Fremling14}, I adopt the $\tau$ gauge: $(A_x,A_y)=B\left(y,-{\tau_1 \over \tau_2}y\right)$. I impose the following periodic boundary conditions:
\be
\label{pbc}
t(L_i)\psi(z,\bar{z})=e^{\i\phi_i}\psi(z,\bar{z}) \quad i=1,2.
\ee
The magnetic translation operator $t$ in the $\tau$ gauge is given by
\be
\label{magnetic translation operator}
t\left(\alpha L_1+\beta L_2\right)=e^{\alpha\partial_1+\beta\partial_2+\i2\pi\beta N_{\phi}\theta_1},
\ee
where $\partial_j\equiv{\partial \over {\partial \theta_j}}$. 

I go over the construction of the CFFS wave function in some detail here. In particular, I show that, in general, there is more than one way to project the CFFS wave function to lowest Landau level (LLL), except for $\nu=1$ and $\nu=1/2$. This allows us to test the sensitivity of the Hall viscosity in the form of the CFFS wave function.

I begin with the Laughlin wave function, which will enter into the construction of the CFFS wave function. In the $\tau$ gauge the unnormalized Laughlin wave function \cite{Haldane85,Fremling16b} for $N$ particles (which will be used in the construction of the CFFS wave function) at filling $\nu=1/m$ can be written as
\begin{widetext}
\be
\label{Laughlin}
\Psi^{\rm Lau}_{m,k}[z_i,\bar{z}_i]=e^{\i\pi\tau N_\phi\sum_i 
\theta_{2,i}^2}
\left[\elliptic{{\phi_1\over 2\pi m}+{k\over m}+{N-1\over2}}{-{\phi_2\over 2\pi}+{m(N-1)\over2}}{mZ \over L_1}{m\tau}\right] 
\prod_{i<j} \left[ \elliptic{{\frac12}}{{\frac12}}{z_{i}-z_{j}\over L_1}{\tau} \right]^m
\ee
where the last factor is the torus analog of the familiar Jastrow factor $\prod_{i<j}(z_i-z_j)^m$ and the factors preceding it ensure the correct boundary conditions and center-of-mass momentum. 
\end{widetext}

Here $Z=\sum_{i}z_{i}$, and $\elliptic[\displaystyle]abz\tau$ is the Jacobi theta functions with rational characteristics, given by\cite{Mumford07}
\be
\elliptic[\displaystyle]abz\tau=\sum_{n=-\infty}^{\infty}e^{\i\pi \left(n+a\right)^2\tau}e^{i2\pi \left(n+a\right)\left(z+b\right)}.
\ee 
The use of these $\vartheta$ functions \cite{Fremling14,Pu20} is advantageous because one does not need to specify the positions of the zeros. Eq.~\ref{Laughlin} has center-of-mass momentum $k=0,1,2\cdots m-1$:
\be
\prod_{i=1}^N t_i(L_1/N_\phi)\Psi^{\rm Lau}_{m,k}[z_i,\bar{z}_i]=e^{\i2\pi({\phi_1\over 2\pi m}+{k\over m}+{N-1\over2})}\Psi^{\rm Lau}_{m,k}[z_i,\bar{z}_i].
\ee

Before projection into the LLL, the CFFS wave function at filling $1/m$ can be written as \cite{Rezayi94}:
\be
\label{unpro}
\Psi^{\rm unpro}_{m,k}[z_i,\bar{z}_i]={\rm Det}\left[e^{i\vec{k}_n\cdot\vec{z}_i}\right]\Psi^{\rm Lau}_{m,k}[z_i,\bar{z}_i]
\ee
Here ${\rm Det}\left[e^{i\vec{k}_n\cdot\vec{z}_i}\right]$ stands for the Slater determinant of the plane waves. To satisfy the periodic boundary conditions Eq.~\ref{pbc}, the allowed wave vectors are 
\be
\vec{k}_n=n_1\vec{b}_1+n_2\vec{b}_2,
\ee
\be
\vec{b}_1=\left({2\pi\over L_1},-{2\pi\tau_1\over L_1\tau_2}\right),
\ee
\be
\vec{b}_2=\left(0,{2\pi\over L_1\tau_2}\right),
\ee
 with both $n_1$ and $n_2$ being integers. Eq.~\ref{unpro} represents a fermionic CFFS if $m$ is even and a bosonic CFFS if $m$ is odd.

Now I need to project the wave function into the LLL. This has been accomplished for the $\nu=1/2$ CFFS using the Jain-Kamilla (JK) projection method \cite{Jain97,Jain97b} in symmetric gauge following Refs.~\cite{Shao15, Geraedts18,Wang19,Pu18}. Here I perform the projection in the $\tau$ gauge for general CFFSs. Let us first consider the fermionic CFFS. The LLL wave function in the $\tau$ gauge can be generally written as $e^{\i\pi\tau N_\phi\sum_i \theta_{2,i}^2}\Psi[z_i]$, in which the second part is a holomorphic function of $z_i$, $i=1,2\cdots N$. The LLL projection is accomplished by using the equation \cite{Fremling19,Girvin84b}
\be
{\PLLL}\left[\bar{z}^ne^{\i\pi \tau N_\phi \theta_2^2}f(z)\right]=e^{\i\pi \tau N_\phi \theta_2^2}\left(z+2\ell^2 \partial_z\right)^nf(z), 
\ee
For the CFFS at $\nu=1/m$, with $m$ being even, I get the LLL projected wave function:
\begin{widetext}
\be
\Psi^{\rm proj}_{m,k}[z_i,\bar{z}_i]=e^{\i\pi\tau N_\phi\sum_i \theta_{2,i}^2}\elliptic{{\phi_1\over 2\pi m}+{k\over m}+{N-1\over2}}{-{\phi_2\over 2\pi}+{m(N-1)\over2}}{m(Z+\i\ell^2 K) \over L_1}{m\tau}{\rm Det}\left[\hat{g}_{nl}\right]\left[ \elliptic{{\frac12}}{{\frac12}}{z_{i}-z_{j}\over L_1}{\tau} \right]^m
\ee
\be
\label{g2}
\hat{g}_{nl}=e^{-\frac{k_n\ell^2}{4}\left(k_n+2\bar{k}_n\right)} e^{\frac{\i}{2}(\bar{k}_n+k_n)z_l}e^{\i k_n\ell^2\partial_{z_l}}
\ee
Here $K$ is the sum over all occupied wave vectors $K=\sum_n k_n$. A determinant of operators ${\rm Det}\left[\hat{g}_{nl}\right]$ is certainly not easy to compute. To avoid this difficulty, one applies the JK projection, and the wave function finally reads:
\be
\label{proj even}
\Psi^{{\rm proj(f)}[\vec{\alpha}]}_{m,k}[z_i,\bar{z}_i]=e^{\i\pi\tau N_\phi\sum_i \theta_{2,i}^2}\elliptic{{\phi_1\over 2\pi m}+{k\over m}+{N-1\over2}}{-{\phi_2\over 2\pi}+{m(N-1)\over2}}{m(Z+\i\ell^2 K) \over L_1}{m\tau}{\rm Det}\left[g^{[\vec{\alpha}]}_{nl}\right]
\ee
\be
\label{ge1}
g^{[\vec{\alpha}]}_{nl}=e^{-\frac{k_n\ell^2}{4}\left(k_n+2\bar{k}_n\right)} e^{\frac{\i}{2}(\bar{k}_n+k_n)z_l}\cdot 
\prod_{p=1}^{m/2}\prod_{j,j\neq l}\elliptic{{\frac12}}{{\frac12}}{z_{l}+\i\alpha_pk_n\ell^2-z_{j}\over L_1}{\tau} 
\ee
Here I have the non-trivial JK projection coefficient $\vec{\alpha}={\rm (\alpha_1, \alpha_2,\cdots \alpha_{m/2})}$, which was first found for $\nu=1/2$ CFFS in Ref.~\cite{Shao15} and for Jain states in Ref.~\cite{Pu17}. For the CFFS wave function to satisfy periodic boundary conditions, they must satisfy $\sum_{p=1}^{m/2} \alpha_p=m$. For $\nu=1/2$, there is only one term, $\alpha_1=2$, and the wave function Eq.~\ref{proj even} is unique. For other fermionic CFFSs with $m\ge 2$, there is, in general, more than one choice of $\alpha_p$. For instance, I can choose $(\alpha_1,\alpha_2)=(2,2),(0,4),(1,3),(5,-1)\cdots$ for $\nu=1/4$. A wave function similar to Eq.~\ref{proj even} was used in Ref.~\cite{Wang19b,Ji19}, which corresponds to the cases $\alpha_{1,2\cdots l}=m/l$ and $\alpha_{l+1,l+2\cdots, m/2}=0$ with $l$ being an integer. The larger class of CFFS wave functions derived above has not been reported before, and it would be interesting to ask in what sense these CFFS wave functions differ.

The JK projection can similarly be applied to bosonic CFFSs with $m\ge 3$. Because $m$ is now an odd number, I factor out a single power of the Jastrow factor. Eq.~\ref{proj even} now becomes
\be
\label{proj odd}
\Psi^{\rm proj(b)[\vec{\alpha}]}_{m,k}[z_i,\bar{z}_i]=e^{\i\pi\tau N_\phi\sum_i \theta_{2,i}^2}\elliptic{{\phi_1\over 2\pi m}+{k\over m}+{N-1\over2}}{-{\phi_2\over 2\pi}+{m(N-1)\over2}}{m(Z+\i\ell^2 K) \over L_1}{m\tau}{\rm Det}\left[g^{[\vec{\alpha}]}_{nl}\right]\prod_{i<j}\elliptic{{\frac12}}{{\frac12}}{z_{i}-z_{j}\over L_1}{\tau}
\ee
\be
\label{go}
g^{[\vec{\alpha}]}_{nl}=e^{-\frac{k_n\ell^2}{4}\left(k_n+2\bar{k}_n\right)} e^{\frac{\i}{2}(\bar{k}_n+k_n)z_l}\cdot 
\prod_{p=1}^{(m-1)/2}\prod_{j,j\neq m}\elliptic{{\frac12}}{{\frac12}}{z_{l}+\i\alpha_pk_n\ell^2-z_{j}\over L_1}{\tau},
\ee
and the condition for $\alpha_p$ becomes: $\sum_{p=1}^{(m-1)/2} \alpha_p=m$.
A special case which is not covered by Eq.~\ref{proj odd} is the bosonic CFFS at filling $\nu=1$. In that case I have only one order of Jastrow factor. One might think that a wave function for the bosonic CFFS for $\nu=1$ can be obtained by dividing the fermionic $\nu=1/2$ CFFS by a single Jastrow factor. It turns out that that wave function is not valid. However, a closely related wave function does the job:
\be
\label{proj 1}
\Psi^{\rm proj(1)}_{m=1,k}[z_i,\bar{z}_i]=e^{\i\pi\tau N_\phi\sum_i \theta_{2,i}^2}\elliptic{{\phi_1\over 2\pi }+{N-1\over2}}{-{\phi_2\over 2\pi}+{(N-1)\over2}}{(Z+\i\ell^2 K) \over L_1}{\tau}{\rm Det}\left[g_{nl}\right]\big/\prod_{i<j}\elliptic{{\frac12}}{{\frac12}}{z_{i}-z_{j}\over L_1}{\tau}
\ee
\be
\label{g1}
g_{nl}=e^{-\frac{k_n\ell^2}{4}\left(k_n+2\bar{k}_n\right)} e^{\frac{\i}{2}(\bar{k}_n+k_n)z_l}\cdot 
\prod_{j,j\neq m}\elliptic{{\frac12}}{{\frac12}}{z_{l}+\i k_n\ell^2-z_{j}\over L_1}{\tau} 
\ee
\end{widetext}
Notice that the argument of the theta function in the preceding equation is slightly different from that of the CFFS for $\nu=1/2$. 

One significant feature for any physical  wave function on a torus is modular covariance. Any physical quantity calculated by these wave functions must be invariant under the transformations $\tau \rightarrow \tau+1$ and $\tau \rightarrow -{1\over \tau}$. More detailed discussions of the modular covariance of composite fermion wave functions were given in Ref.~\cite{Pu20}. In Appendix~\ref{modular covariance}, I prove the general CFFS wave functions are modular covariant and show numerical confirmations. Another issue is how close the JK projected wave functions with different $\vec{\alpha}$ at the same filling factor are. In Appendix~\ref{JK projection}, I show that the JK projected CFFS wave functions have very high overlap with the directly projected wave function and also with each other at the same fillings.

\section{Results and Discussion}
\label{results}

I evaluated the Hall viscosities  for the CFFS wave functions at $\nu=1$, 1/2, 1/3, 1/4, 1/5 and 1/6,  
using the Berry curvature expression given in Eq.~\ref{berry-curv}. The results are summarized in Table~\ref{table1}, along with the CFFS shapes I used. The CFFS shapes correspond to the global ground states for the given system sizes.  I adopt the standard variational Monte Carlo method for unprojected wave functions, and the lattice Monte Carlo method \cite{Wang19} for LLL projected wave functions. The last digit in the parentheses in Table~\ref{table1} represents the statistical error. I present $\eta^A$ in units of $\hbar \rho/4$, where $\rho=N/V$ is the particle density. In these units, $\eta^A$ is given by the shift $\sh$ for gapped FQH states according to Read's relation Eq.~\ref{hall visc}. All results in Table~\ref{table1} are evaluated for a square torus. As shown below, the results are quite stable as the system size increases, as long as the CFFS shapes stay circular. 

As I mentioned, there are two key issues here: 1) What is the value of Hall viscosities for CFFSs? 2) Are they still topologically quantized? The fermionic CFFS at filling $\nu=1/2p$ can be regarded as  the $m\rightarrow \infty$ limit of Jain states at filling $\nu={m\over 2pm\pm 1}$. Ref.~\cite{Pu20} shows that the Hall viscosities for Jain states are $\eta^A=(\pm m+2p){\hbar \rho \over 4}$. Taking the limit, the Hall viscosity for the CFFS diverges to $+\infty$ or $-\infty$ depending on whether I approach it from below or above. Meanwhile, Table~\ref{table1} shows that the Hall viscosities are finite and change very little with system size, although in general they are not quantized at integer values in units of ${\hbar \rho \over 4}$.  
 
 Among these cases, the CFFS at $\nu=1/2$ has an additional symmetry, namely, an exact particle-hole symmetry for any two-body interaction confined in the LLL. The CFFS wave function satisfies this symmetry to a high degree but not exactly. Refs.~\cite{Read10,Haldane09} show that the particle-hole symmetric ground state has $\eta^A={\hbar \rho\over 4}$, independent of whether the ground state is compressible or incompressible. Table~\ref{table1} shows that the Hall viscosity for the LLL projected CFFS wave function at filling $1/2$ is quantized at $\hbar \rho\over 4$ for a square torus with the same accuracy as found for the incompressible Jain states \cite{Pu20}. This shows that the small particle-hole asymmetric part in the CFFS wave function does not change the Hall viscosity appreciably for $\tau=\i$. Levin and Son \cite{Levin17} showed that if the particle-hole symmetry is not spontaneously broken, there is an exact relationship between the Hall conductivity and the susceptibility which can be derived through Dirac CF theory. This relation predicts the orbital spin of the $\nu=1/2$ CFFS to be $s=1/2$ by making use of Galilean invariance \cite{Hoyos12,Bradlyn12}. The orbital spin is half of the shift. Levin and Son also showed that from Halperin-Lee-Read theory one can derive a similar relation which predicts $s=1$ for $\nu=1/2$ CFFS. My result is consistent with orbital spin predicted by the Dirac CF theory \emph{provided} that the Hall viscosity and orbital spin for the CFFS are also related through Read's conjecture.   

What about the Hall viscosities of CFFSs at other fillings which do not have particle-hole symmetry? Ref.~\cite{Milovanovic10} argued that the Hall viscosity 
of $\nu=1/m$ CFFS is $\eta^A={m\hbar\rho\over 4}$, \ie the same value as Laughlin states for all fillings (even without particle-hole symmetry) in the thermodynamic limit if it is evaluated through Eq.~\ref{berry-curv} at $\tau=\i$. My numerical result is  close to that prediction only for the bosonic CFFS wave function at $\nu=1$ in LLL. I found that the Hall viscosities are not quantized at integer values of $\sh$ through Eq.~\ref{hall visc} in general. In fact, they are more or less around (not accurately at) the values $\eta^A={m\rho\hbar\over 8}$ for $\nu=1/m$, which corresponds to $\sh=m/2$ (except for the $\nu=1$ case just mentioned). Ref.~\cite{Goldman18} gives a conjecture for the Hall viscosities of CFFS which says $\eta^A={\hbar\rho\over 4}(m-1)$ for $\nu=1/m$. For $\nu=1/2$ it corresponds to the orbital spin of a Dirac composite fermion and also agrees with my numerical results.

\begin{table*}
\makebox[\textwidth]{\begin{tabular}{|c|c|c|c|c|c|}
\hline
\multicolumn{2}{|c|}{Shape of CFFS}&\includegraphics[width=0.1\textwidth, height=0.1\textwidth]{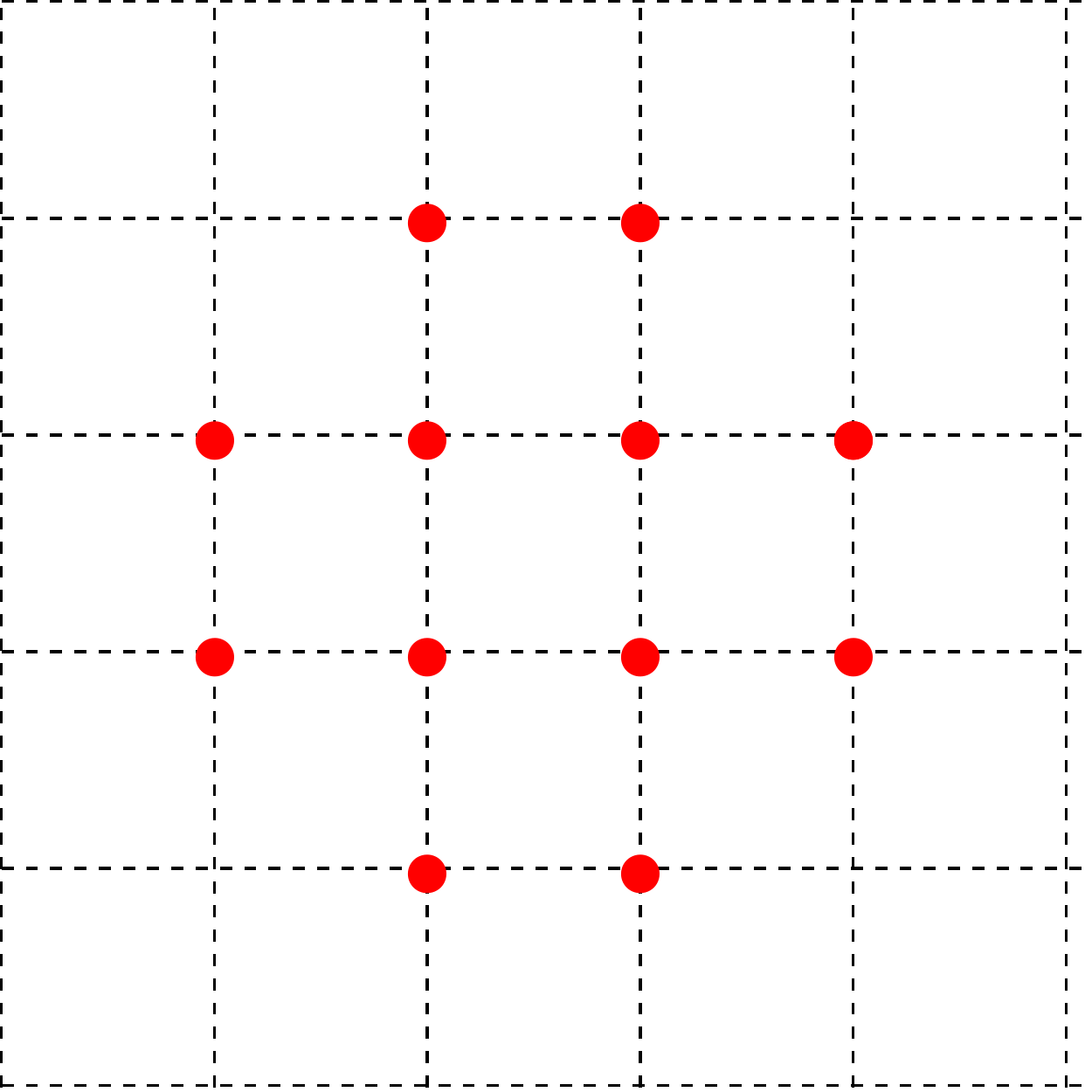}&\includegraphics[width=0.1\textwidth, height=0.1\textwidth]{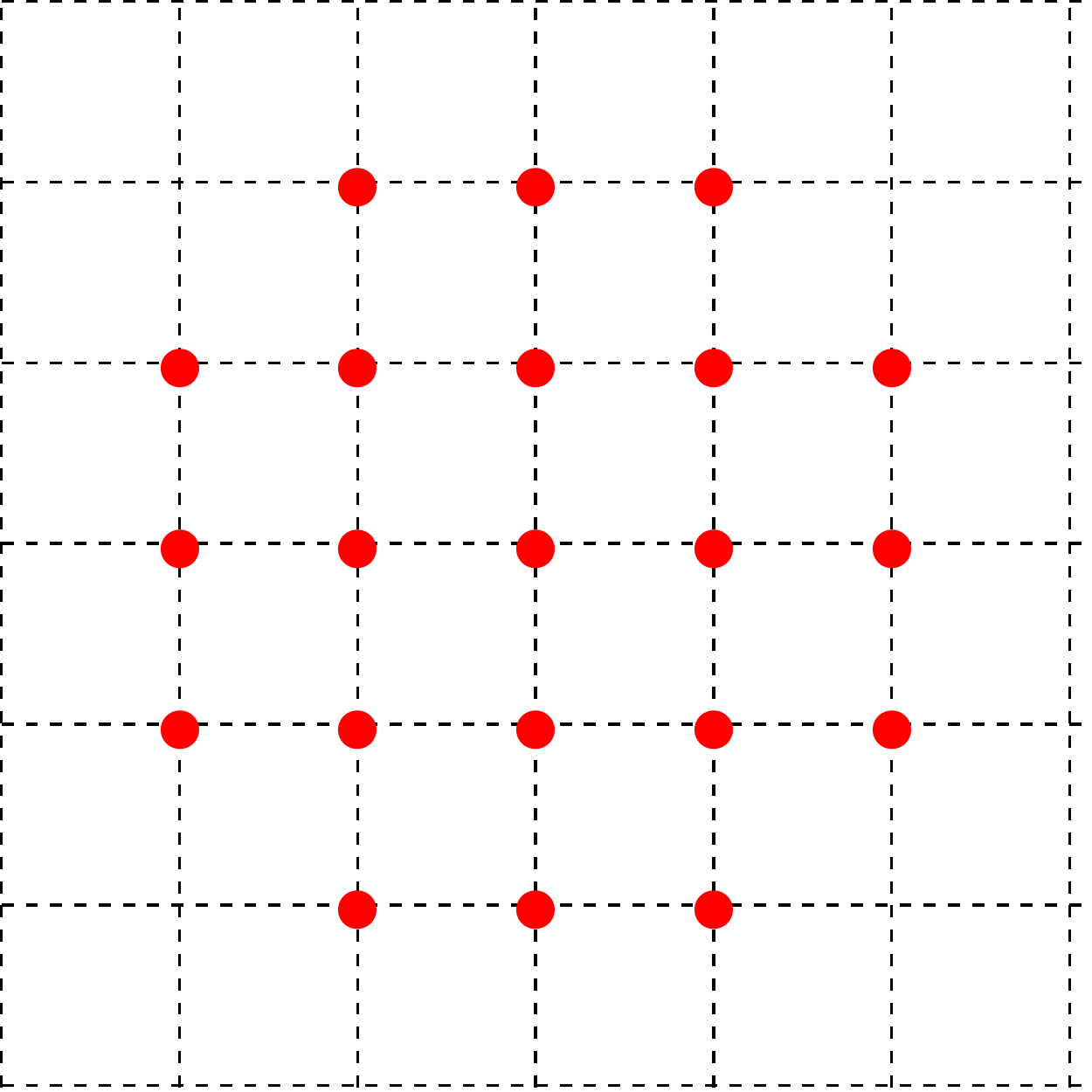}&\includegraphics[width=0.1\textwidth, height=0.1\textwidth]{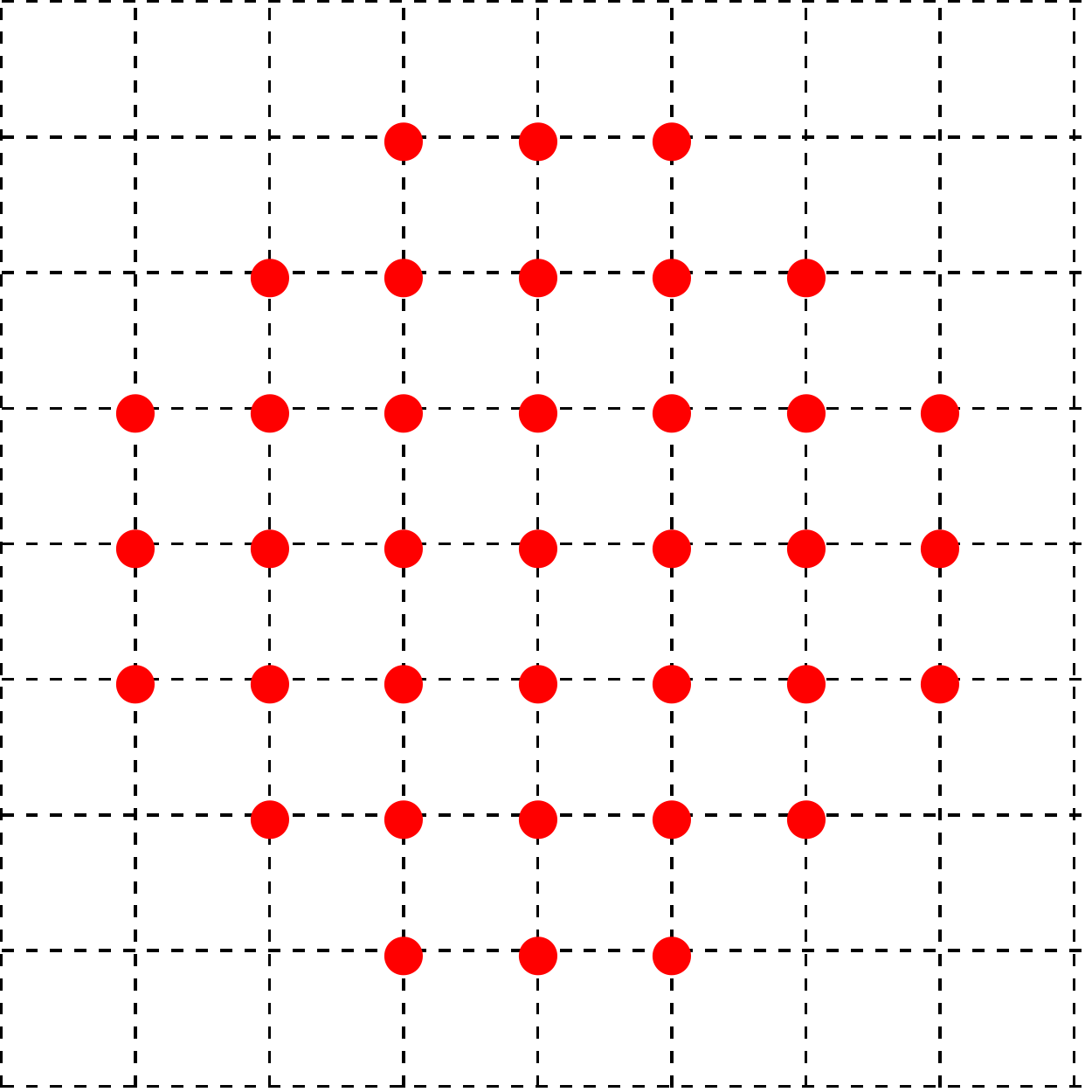}&\includegraphics[width=0.1\textwidth, height=0.1\textwidth]{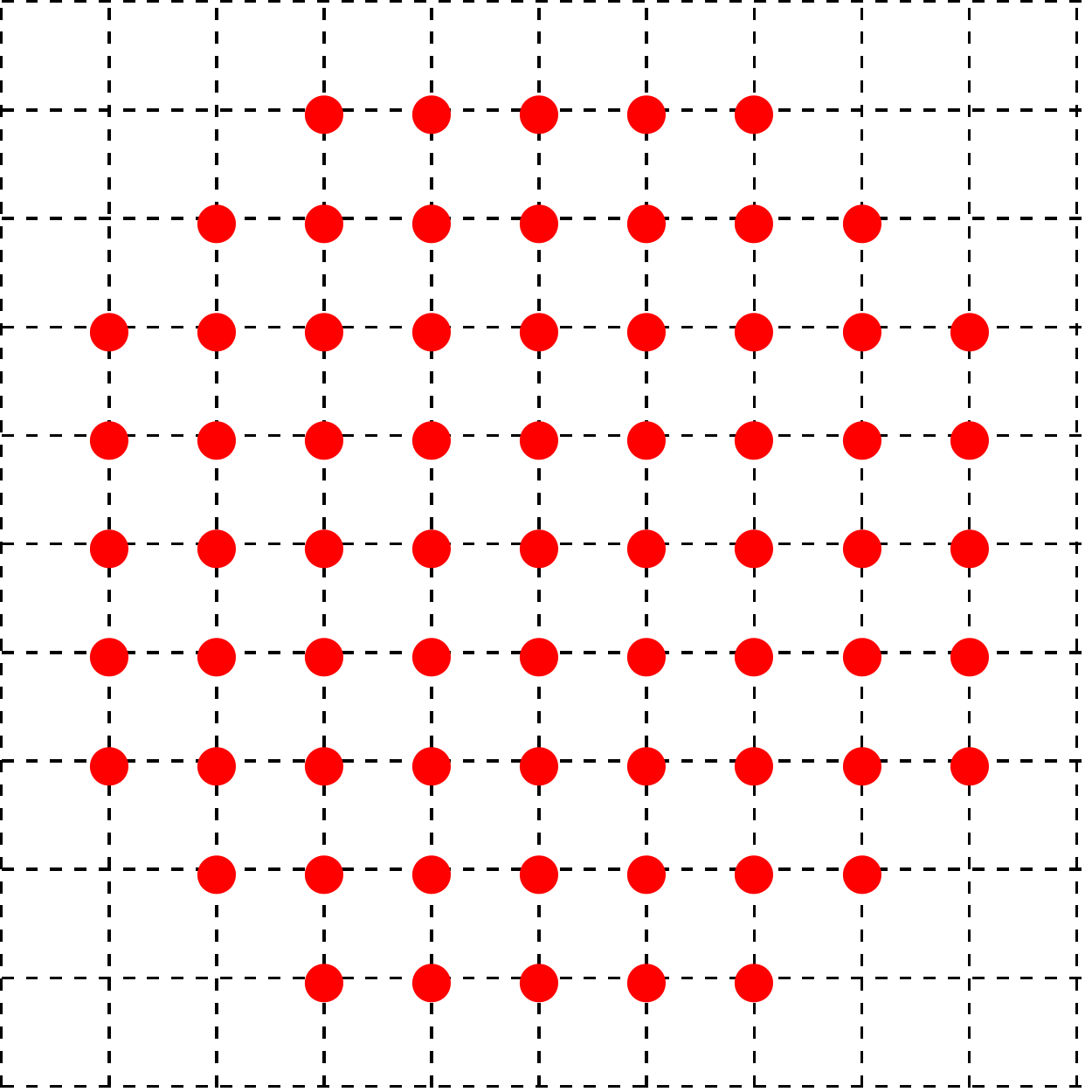}\\ \hline
\multicolumn{2}{|c|}{Number of particles}&12&21&37&69\\ \hline
$\nu$&$\vec{\alpha}$&\multicolumn{4}{|c|}{$4\eta^A/\hbar \rho$}\\ \hline
\multirow{2}{*}{$1/2$}&(2)&1.013(1)&1.012(4)&1.014(1)&1.011(4)\\ \cline{2-6}
&un&1.054(1)&1.040(3)&1.058(6)&1.05(1)\\  \hhline{|=|=|=|=|=|=|}
\multirow{3}{*}{$1/4$}&(2,2)&1.991(3)&2.06(1)&2.087(4)&2.03(1)\\ \cline{2-6}
&(4,0)&1.801(2)&1.80(1)&1.822(5)&1.783(6)\\ \cline{2-6}
&un&2.339(3)&2.327(6)&2.38(2)&2.33(2)\\  \hhline{|=|=|=|=|=|=|}
\multirow{4}{*}{$1/6$}&(2,2,2)&3.355(7)&3.402(9)&3.47(2)&3.31(8)\\ \cline{2-6}
&(3,3,0)&3.147(6)&3.203(8)&3.28(3)&3.11(4)\\ \cline{2-6}
&(6,0,0)&2.929(4)&2.92(2)&2.99(2)&2.82(3)\\ \cline{2-6}
&un&3.712(6)&3.67(1)&3.78(2)&3.57(3)\\  \hhline{|=|=|=|=|=|=|}
\multirow{2}{*}{$1$}&(1)&1.041(3)&1.115(4)&1.152(3)&1.170(5)\\ \cline{2-6}
&un&0.4700(7)&0.4509(7)&0.452(2)&0.449(5)\\  \hhline{|=|=|=|=|=|=|}
\multirow{2}{*}{$1/3$}&(3)&1.336(1)&1.336(7)&1.346(3)&1.327(7)\\ \cline{2-6}
&un&1.681(2)&1.670(4)&1.708(7)&1.69(2)\\  \hhline{|=|=|=|=|=|=|}
\multirow{3}{*}{$1/5$}&(3,2)&2.533(5)&2.593(5)&2.64(1)&2.54(2)\\ \cline{2-6}
&(5,0)&2.340(4)&2.35(1)&2.379(5)&2.28(1)\\ \cline{2-6}
&un&3.015(3)&3.00(1)&3.08(2)&2.97(2)\\ \hline
\end{tabular}}
\caption{\label{table1} The Hall viscosity for CFFS wave functions. The data are shown as $4\eta^A/\hbar \rho$, where $\eta^A$ defined in Eq.~\ref{berry-curv} is calculated through Monte Carlo at $\tau=\i$ and $\rho$ is the particle density $\rho=N/V$. For gapped FQH states, the quantity shown would be equivalent to shift $\sh$. I specify the CFFS shapes (which correspond to the global ground states) and system sizes Eq.~\ref{berry-curv} in the first two rows. I specify the wave function at each filling factor by showing $\vec{\alpha}$, which is defined in Sec.~\ref{wf}. The unprojected wave function is denoted as``un''. The last digit in parentheses represents the statistical error.}
\end{table*}

\begin{table*}
\makebox[\textwidth]{\begin{tabular}{|c|c|c|c|c|}
\hline
\multicolumn{2}{|c|}{Shape of CFFS}&\includegraphics[width=0.08\textwidth, height=0.138\textwidth]{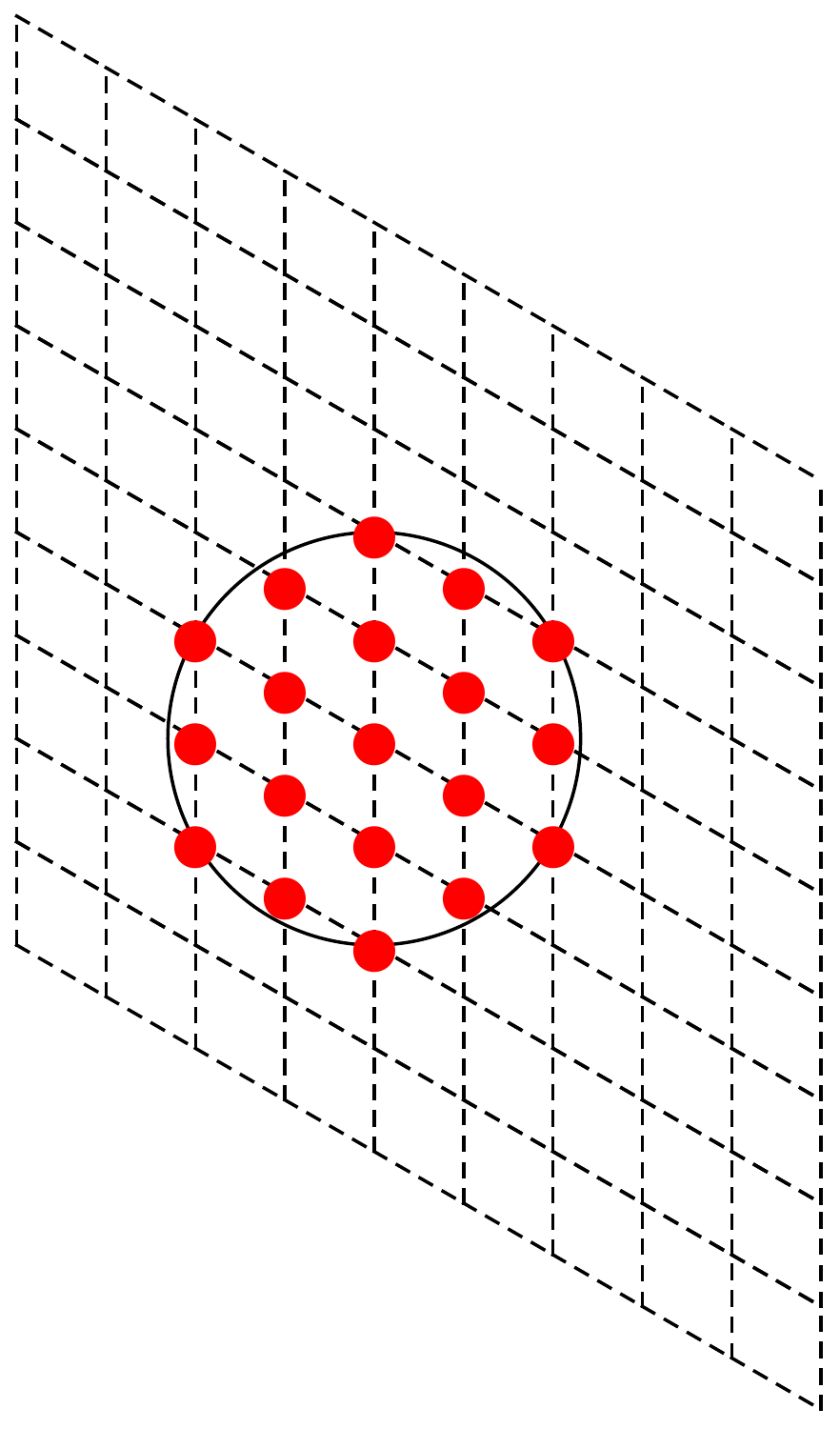}&\includegraphics[width=0.08\textwidth, height=0.138\textwidth]{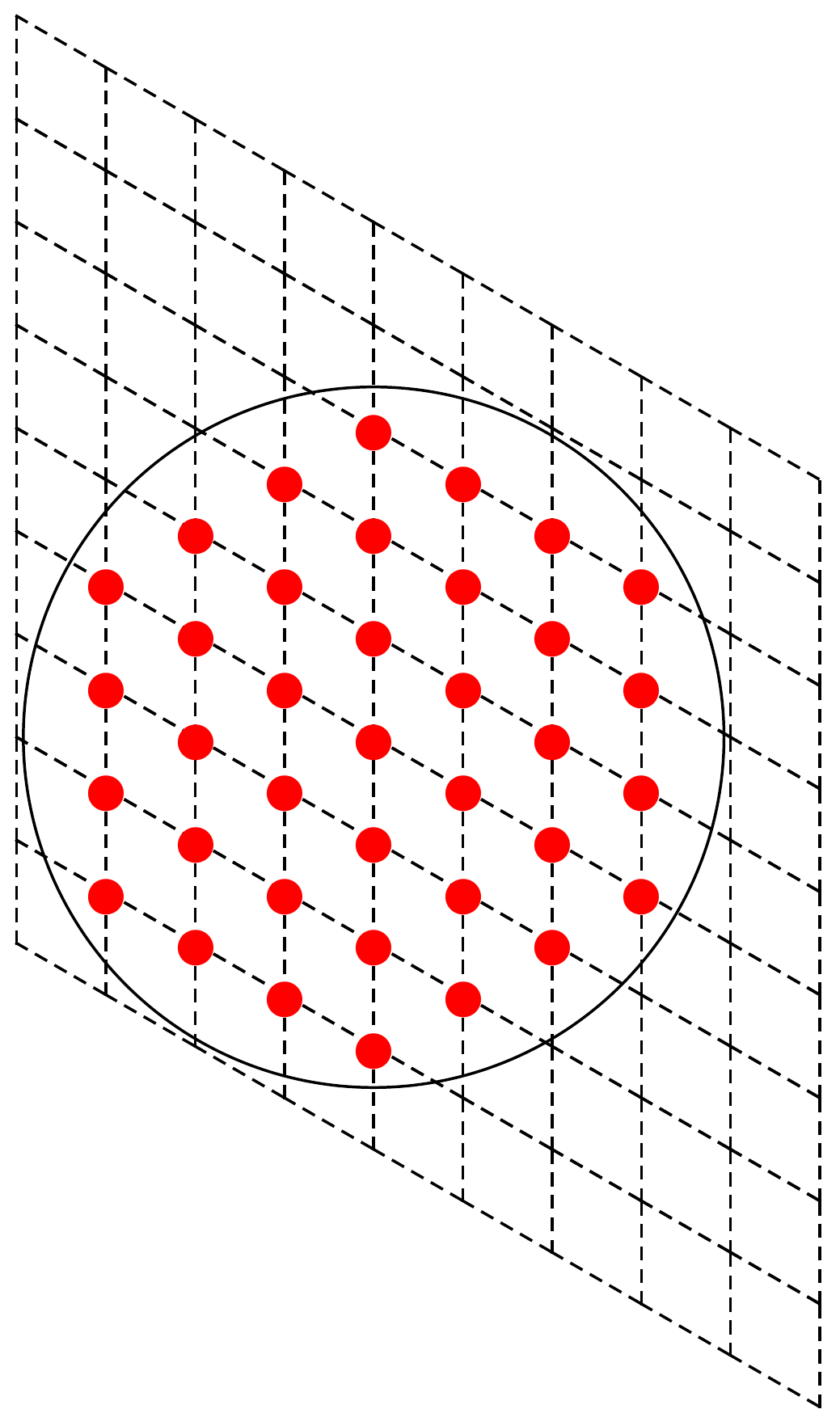}&\includegraphics[width=0.08\textwidth, height=0.193\textwidth]{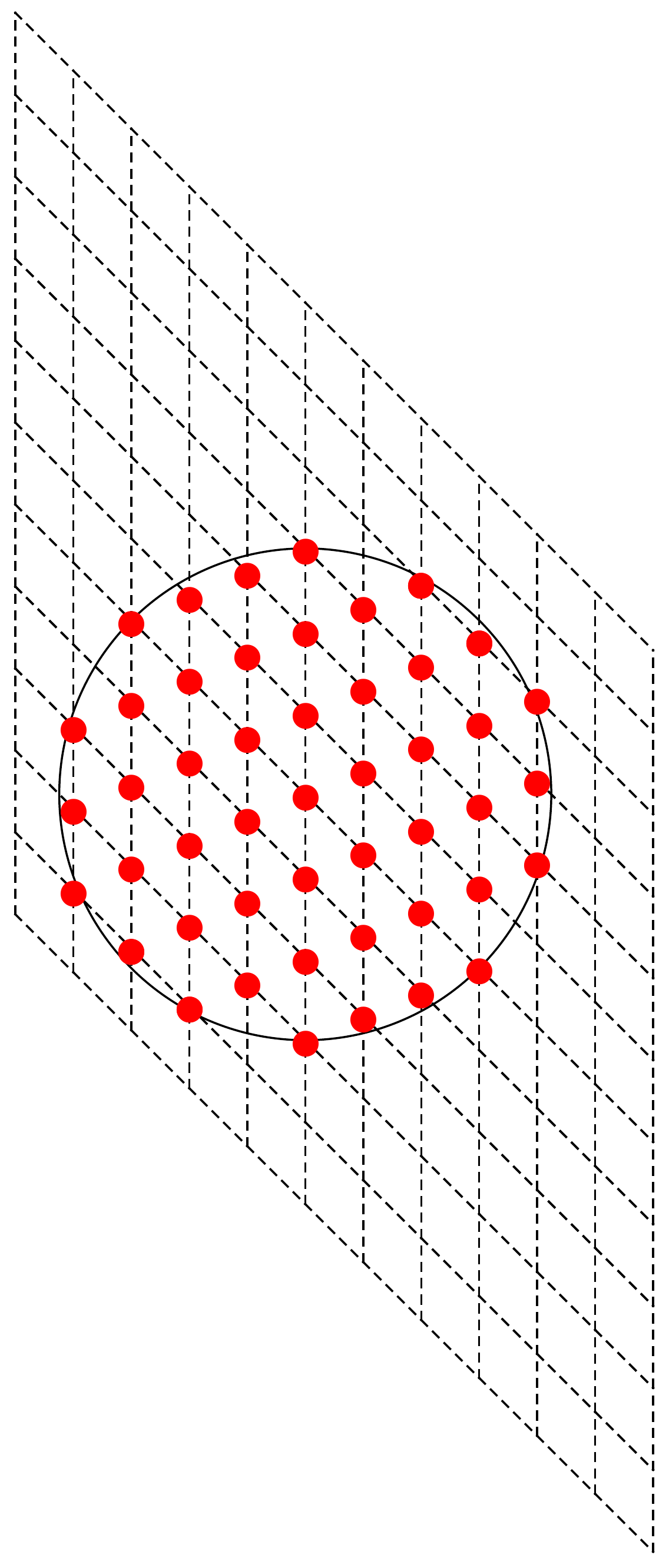}\\ \hline
\multicolumn{2}{|c|}{Number of particles}&19&37&47\\ \hline
\multicolumn{2}{|c|}{$\tau$}&\multicolumn{2}{|c|}{$e^{\i\pi/3}$}&$e^{\i\pi/4}$\\ \hline
$\nu$&$\vec{\alpha}$&\multicolumn{3}{|c|}{$4\eta^A/\hbar \rho$}\\ \hline
$1/2$&(2)&1.011(2)&1.014(2)&0.945(5)\\ \hline
$1/4$&(2,2)&2.04(1)&2.073(7)&1.533(7)\\ \hline
$1/6$&(6,0,0)&2.89(2)&2.97(2)&1.13(2)\\ \hline
$1$&(1)&1.084(2)&1.144(3)&1.151(2)\\ \hline
$1/3$&(3)&1.333(4)&1.339(6)&1.116(3)\\ \hline
$1/5$&(3,2)&2.57(1)&2.62(2)&1.57(2)\\ \hline
\end{tabular}}
\caption{\label{table2} The Hall viscosity for CFFS wave functions for different shapes of the torus using the same convention as in Table~\ref{table1}. The first two CFFS shapes have hexagonal torus shapes, while the third one has $\tau=e^{i\pi/4}$. I choose the CFFS shape to be as circular as possible for each torus so that the CFFS wave functions correspond to the global ground states. Note that the $N=37$ column has a different wave vector configuration than the $N=37$ column in Table~\ref{table1}. The last digit in parentheses represents the statistical error.}
\end{table*}

The second question is whether the Hall viscosity calculated through Eq.~\ref{berry-curv} is still topologically quantized for CFFSs, just like the gapped FQH states. I found the answer is no. The answer can be seen from two aspects. First, Table~\ref{table1} shows that the Hall viscosities are different for different CFFS wave functions at the same filling; that is, the Hall viscosity depends on whether the wave functions are projected into the LLL or not, and if so, how they are projected.  However, different CFFS wave functions at a given filling actually have very high overlap as shown in Appendix~\ref{JK projection} and describe the same topological phase, so they should yield the same value if the Hall viscosity is topologically quantized. Indeed, for the Jain states Ref.~\cite{Pu20} found that the projected and unprojected CF wave functions have the same Hall viscosities if the system is big enough. 
Second, Refs.~\cite{Read10,Pu20} have shown that the topologically quantized Hall viscosities would yield the same value independent of at which $\tau$ point the Berry curvature is evaluated; that is, it is independent of the shape of the torus. I test whether this is still true for CFFSs, and the results are summarized in Table~\ref{table2}. For a gapped FQH state, two ground states at different $\tau$ values are adiabatically connected, in the sense that I can get one from the other by just changing $\tau$ in the wave function. This is apparently not true for the CFFS. When the shape of the torus is deformed, the shape of the CFFS also changes. A circular CFFS at $\tau$ would become an elliptical CFFS at $\tau'$. While the elliptical CFFS might still have the most compact shape in that given momentum sector, it is no longer the global ground state for all momentum sectors. As shown in Ref.~\cite{Fremling18}, the ground states in high-energy momentum sectors are not very accurately described by the CFFS wave functions. Therefore, I changed the CFFS configuration in $k$ space to make the CFFS as circular as possible for $\tau=e^{\i \pi/3},e^{\i \pi/4}$ considered here. As Table~\ref{table2} shows, the Hall viscosity for $\tau=e^{\i \pi/3}$ is very close to the Hall viscosity for $\tau=i$ shown in Table~\ref{table1}, which is remarkable because they are two different ground states that are not adiabatically connected. However, the Hall viscosity for $\tau=e^{\i \pi/4}$ clearly deviates from the values of a square torus. Hence, I conclude that the Hall viscosity of the CFFS is not topologically quantized in the $\tau$ space.

Ref.~\cite{Pu20} presents a way to analytically derive the Hall viscosity for the CF wave functions. That approach is based on the assumption that the overall normalization factor does not contribute to Hall viscosity. For the CFFS wave function in Eq.~\ref{unpro}, the normalization of the Slater determinant is independent of $\tau$ because $e^{\i \vec{k}\cdot\vec{z}}=e^{\i 2\pi\left(n_1\theta_1+n_2\theta_2\right)}$ if $\vec{k}=n_1\vec{b}_1+n_2\vec{b}_2$ and $z=\theta_1L+\theta_2L\tau$. Therefore, the aforementioned assumption for CFFS means
\be
\lim_{N\rightarrow \infty} {1\over N} \left({\partial\over \partial \tau_2}\right)_\tau\ln Z= 0.
\label{Z_infty}
\ee
\be
\lim_{N\rightarrow \infty} {1\over N} \left({\partial\over \partial \tau_2}\right)_\tau\ln Z\begin{cases}
=0 &\text{Jain states}\\
\neq 0&\text{CFFS}\end{cases}.
\ee
\be
Z=\left({\int \prod_id^2r_i |\Psi_{m,k}^{\rm unpro}|^2\over \int \prod_id^2r_i |\Psi_{m,k}^{\rm Lau}|^2}\right)^{-{1\over 2}}
\ee
If the assumption in Eq.~\ref{Z_infty} is also true for the CFFS, then the Hall viscosity of the CFFS would be $\eta^A=m{\hbar \rho\over 4}$ at $\nu=1/m$ following the approach in Ref.~\cite{Pu20}. The results in Table~\ref{table1} clearly show that this is not the case. It is interesting to note that while the overall normalization factor does not contribute to the Hall viscosity for gapped Jain states, it makes a nontrivial contribution to the Hall viscosity of the CFFSs.

\section{Conclusion}

In this paper, I calculated the Hall viscosity of CFFS wave functions at various fillings by evaluating the Berry curvature in the $\tau$ space. I found that the Hall viscosities at $\tau=\i$ are finite and fluctuate very little with system size. Especially, the CFFS wave function at filling $1/2$ in LLL has a Hall viscosity very close to $\eta^A={\hbar \rho\over 4}$, as expected for a particle-symmetric state \cite{Read10}. This value is also consistent with the $1/2$ orbital spin of Dirac composite fermions \cite{Levin17}. I also compared my numerical results with more general theoretical conjectures \cite{Milovanovic10,Goldman18} for other fillings but have not found a perfect consistency. I also note that the Hall viscosities depend on whether the CFFS wave functions are projected or not and the method of projection. Even though the JK projected wave functions have very high overlaps at the same filling factors, the Hall viscosities are not the same. Similar sensitivity to the exact form of the wave functions was found for the Berry curvature in Ref.~\cite{Ji19}. By evaluating the Hall viscosity at different positions in $\tau$ space, I show the Hall viscosity defined through the Berry curvature of the CFFS is not topologically quantized. 

My work leaves some questions that need further elucidation: Is there a general relation between Hall viscosity of the CFFS at $\tau=i$ and the shift (or orbital spin), as my result for $\nu=1/2$ indicates? If the answer is yes, how can one interpret the values found in Table~\ref{table1}, and why do the different LLL projections yield different values? Are some flux attachments ({\em i.e.}, $\vec{\alpha}$ in Table~\ref{table1}) more physical than others? I hope these problems will be addressed in the future.

Another important question is whether the Hall viscosity of CFFSs is manifested in any physical quantity in an experiment. Ref.~\cite{Hoyos12,Bradlyn12} showed that for gapped FQH states the Hall viscosity shows up in the electromagnetic response by imposing the nonrelativistic diffeomorphism invariance on the low-energy effective theory. To my knowledge, how this works for the CFFSs remains an open question. Meanwhile, my numerical results suggest that the Hall viscosity of CFFSs is not a universal quantity in general. Nonetheless, all different wave functions give similar values, which gives me confidence that the Hall viscosity of CFFSs is captured by this calculation.

In an interesting study, Ref.~\cite{Liu17} proposed a generalized definition for the dissipationless viscosity when studying the phonon dynamics of magnetic metals. The dissipationless viscosity is defined as part of the Green's function that determines the electron-ion force. Ref.~\cite{Liu17} and its Supplemental Material showed that for insulators, their generalized definition is equivalent to the Berry curvature format, while it reduced to a different form for metals. It would be interesting to find out how their generalized definition applies to the Hall viscosity of CFFSs, and whether the results would be different from the results shown in this work.

\begin{acknowledgments} 
I am grateful to J. K. Jain for our discussion on the results and his advice on the manuscript. I also would like to thank Michael Fremling for his comments on the manuscript and D. T. Son for our discussion that inspired this work. This work was supported in part by the U. S. Department of Energy, Office of Basic Energy Sciences, under Grant No. DE-SC0005042. The numerical part of this research was conducted with Advanced CyberInfrastructure computational resources provided by the Institute for CyberScience at the Pennsylvania State University. I am grateful to DIAMHAM for the performance of exact diagonalizations.
\end{acknowledgments}
\begin{appendix}
\section{modular covariance of the wave function}
\label{modular covariance}
As Refs.~\cite{Fremling19,Pu20} have proved, the Jain wave functions for FQH states are modular covariant. The CFFS wave functions considered in this paper are also modular covariant. The reasoning is simple. The unprojected wave function Eq.~\ref{unpro} is made up of two parts, and $\Psi^{\rm Lau}$ has been shown to be modular covariant \cite{Fremling19}. The plane wave part can be written as ${\rm Det}\left[e^{i\vec{k}_n\cdot\vec{z}_i}\right]$. Since both $\vec{k}_n$ and $\vec{z}_m$ are invariant under modular transformations, the plane wave part is modular invariant. Thereby, the unprojected CFFS wave function is modular covariant. And as Ref.~\cite{Fremling19} proved, the LLL projection preserves the modular covariance.

\begin{figure}[t]
	\includegraphics[width=\columnwidth]{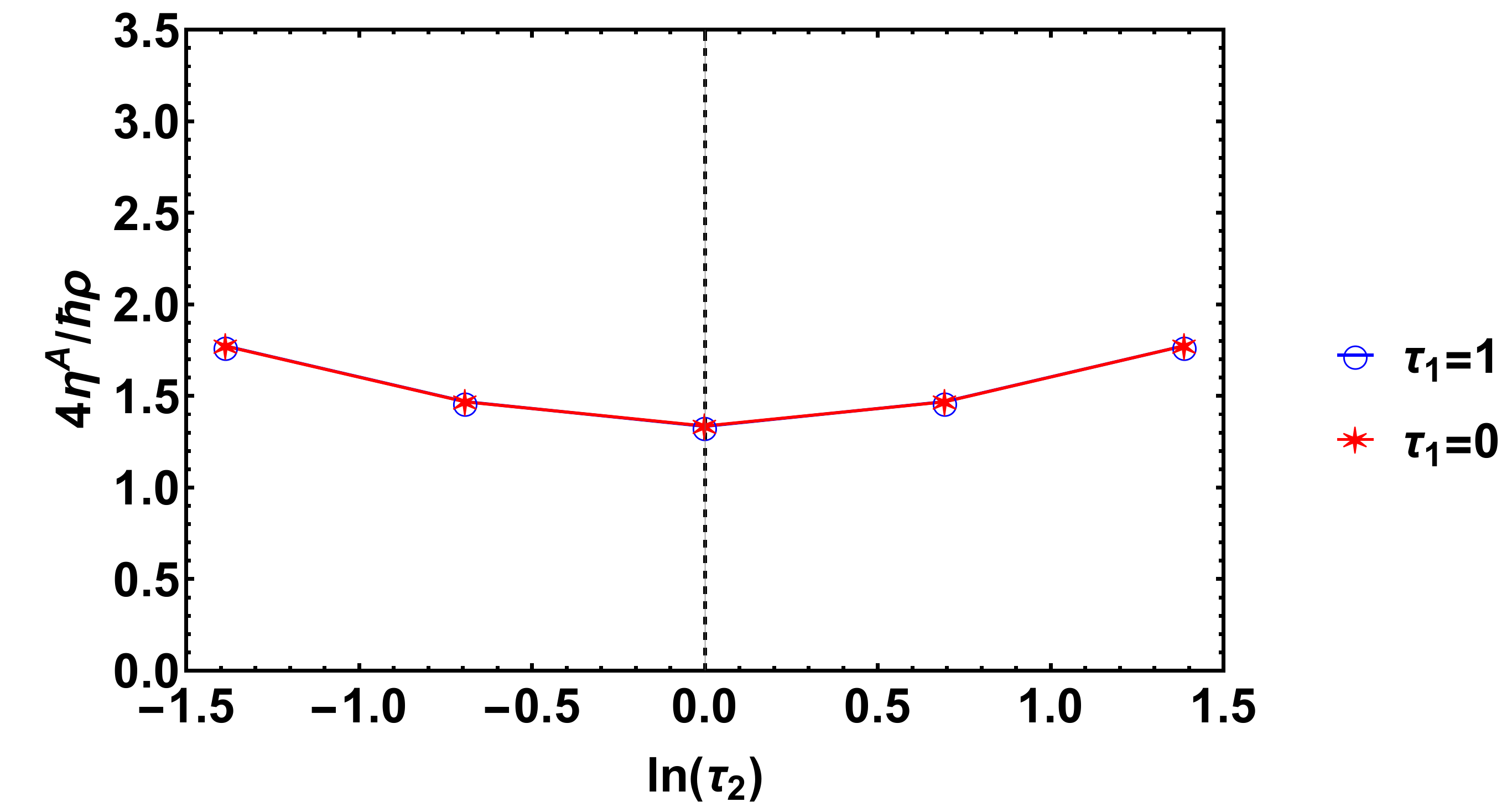} 
	\caption{
 The Hall viscosity for the 12-particle $\nu=1/3$ bosonic CFFS state at several positions in the $\tau$ plane.
 The different symbols represent different values of $\tau_1$,
 whereas the horizontal axis plots $\ln(\tau_2)$. The red curve is symmetric about the dashed line, which is consistent with the symmetry under $\tau\rightarrow -1/\tau$ transformation. The red and blue curves overlap, which is consistent with the symmetry under $\tau\rightarrow \tau+1$ transformation.
 }
	\label{fig2}
\end{figure}

Fig.~\ref{fig2} shows the Hall viscosities of the $\nu=1/3$ bosonic CFFS wave function for different $\tau$'s that are connected by modular transformations. First, the red curve is symmetric about the dashed line, which is consistent with the symmetry under the $\tau\rightarrow -1/\tau$ transformation. Second, the red and blue curves overlap, which is a result of the symmetry under the $\tau\rightarrow \tau+1$ transformation.

\section{Validity of Jain-Kamilla projection for general CFFS wave functions}
\label{JK projection}
While the projected wave function generated by JK projection is not exactly the same as the wave function generated by direct projection, it is an accurate approximation of the latter. This was confirmed for Jain states and CFFS at $\nu=1/2$ in torus geometry by comparing the energies and overlaps between JK projected wave functions and the Coulomb ground states \cite{Pu17,Pu18}. To confirm the validity of JK projection for general CFFS wave functions, I adopt two approaches.

In the first approach, I calculate the overlaps between the JK projected wave functions and the directly projected wave function. Since the system size accessible for direct projection is very limited, here I show only the results of $\nu=1/4$ CFFS of a four-particle system. As shown in Table.~\ref{table3}, the overlaps between the directly projected wave function and the two different JK projected wave functions $(2,2)$ and $(4,0)$ are both very close to unity, which suggests the validity of JK projection.

In the second approach, I evaluate the overlaps between the JK projected wave functions with different $\vec{\alpha}$s. The results are summarized in Table.~\ref{table4}. The overlaps are all nearly unity, which is self-consistent evidence that they are all accurate approximations of the directly projected wave functions. I calculate these overlaps for systems on a square torus with the shapes of CFFSs shown in Table~\ref{table1} and for systems on a nearly square torus with the same shapes of CFFSs. Given the high overlaps between JK projected wave functions with different $\vec{\alpha}$ for the square and nearly square tori, it is clear that the differences of Hall viscosities shown in Table~\ref{table1} are caused by the tiny detailed differences between JK projected wave functions and that the Hall viscosity of CFFS is not a topological quantity in general, in contrast to the gapped FQH states.

\begin{table*}
\makebox[\textwidth]{\begin{tabular}{|c|c|}
\hline
$|\langle \Psi_{\rm Dir}|(2,2)\rangle|$&$|\langle \Psi_{\rm Dir}|(4,0)\rangle|$\\ \hline
$0.99572(6)$&$0.99963(1)$\\ \hline
\end{tabular}}
\caption{\label{table3} The amplitudes of overlaps between the directly projected wave function and JK projected wave functions for the $\nu=1/4$ CFFS with four particles (two by two) on a square torus. The $\Psi_{\rm Dir}$ is obtained by directly projecting Eq.~\ref{unpro} to the lowest Landau level. The JK projected wave function is shown in Eq.~\ref{proj even} and labeled by $\vec{\alpha}$. The last digit in parentheses represents the statistical error.}
\end{table*}

\begin{table*}
\makebox[\textwidth]{\begin{tabular}{|c|c|c|c|}
\hline
$\nu$&$|\langle\vec{\alpha}|\vec{\alpha}'\rangle|$&Overlap amplitude $\tau=i$&Overlap amplitude $\tau=0.99i$\\ \hline
$1/4$&$|\langle (4,0)|(2,2)\rangle|$&$0.9911(1)$&$0.9911(1)$\\ \hline
$1/5$&$|\langle (5,0)|(3,2)\rangle|$&$0.9944(1)$&$0.9944(1)$\\ \hline
\multirow{3}{*}{$1/6$}&$|\langle (2,2,2)|(6,0,0)\rangle|$&$0.9946(1)$&$0.9946(2)$\\ \cline{2-4}
&$|\langle (3,3,0)|(6,0,0)\rangle|$&$0.9955(1)$&$0.9954(2)$\\  \cline{2-4}
&$|\langle (3,3,0)|(2,2,2)\rangle|$&$0.99962(2)$&$0.99962(2)$\\  \hline
\end{tabular}}
\caption{\label{table4} The overlaps between different JK projected states labeled by $\vec{\alpha}$. The particle number is 12 and the shape of the CFFS is shown in Table~\ref{table1} (on a square torus). The third column is for an exact square torus, and the fourth column is for a rectangular torus which is very close to a square torus. The last digit in parentheses represents the statistical error.}
\end{table*}

\end{appendix}

\bibliographystyle{apsrev}

\end{document}